\documentstyle[12pt] {article}

\newcommand{\beq}{\begin{equation}}
\newcommand{\eeq}{\end{equation}}
\newcommand{\beqa}{\begin{eqnarray}}
\newcommand{\eeqa}{\end{eqnarray}}
\newcommand{\ba}{\begin{array}}
\newcommand{\ea}{\end{array}}

\begin{document}

\begin{center}
\large
{\bf Turbulence and Bifurcation in\\
the Motion of an Hydrocyclone} 

\vskip 1. truecm

\normalsize
{\bf Maria Morandi Cecchi}$^{(1)}$ and {\bf Luca Salasnich}$^{(1)(2)}$
\vskip 0.5 truecm
$^{(1)}$ Dipartimento di Matematica Pura ed Applicata, 
Universit\`a di Padova, Via Belzoni 7, I--35131 Padova, Italy \\
$^{(2)}$ Istituto Nazionale per la Fisica della Materia, Unit\`a di Milano, \\
Via Celoria 16, I--20133 Milano, Italy
\end{center}

\vskip 0.5 truecm

\section{The Hydrocyclone}

The hydrocyclone is a tool used in different fields but mostly in mining 
industries to separate minerals. 
The minerals are put in a liquid or in a stable suspension of a chosen density 
which must be less than the density of the heavier minerals and 
greater than that of the lighter minerals [1,2,3]. 
\par
Nowadays there is a strong interest for such tools, mostly in the 
processes where is possible to eliminate or reduce a lot the operation 
of grinding. This allows to reduce the cost of production and simplify 
the separation of minerals [4,5,6,7]. 
The modes of separation are two. One is the {\it static mode}, 
where the basis of the separation process is the gravitational force. 
The other is the {\it dynamic mode}, which operates by using a 
centrifugal force. 
\par
The static mode can be applied when is valid 
the relation $F_g > F_D$, where $F_g$ is the gravitational force and 
$F_D$ is the resistance force to the motion of mineral particles 
into the dense fluid. 
The relation is strictly related to the mass and dimension of the particles. 
As a consequence, the static separation can be used only with dimensions 
of the particles greater than $4$--$6$ mm. Instead the dynamic separation 
can be used with a dimension of the particles less than $0.2$--$0.3$ mm. 
\par
The hydrocyclone can be classified as: \\
i) conic hydrocyclone (DMS); \\
ii) cylindric hydrocyclone (DWP); \\
iii) multi--stage hydrocyclone (TRI--FLOW). 
\par
The conic hydrocyclone is a cylinder connected in the lower part with a 
truncated cone of an angle between $15^o$--$20^o$. The discharge of materials 
with less density (overflow) is done through a tube, the so--called 
{\it vortex finder}, with center in the upper cylinder. 
The discharge of the heavier materials 
(underflow) is realized through a tube ({\it apex}) 
which is connected to the lower base of the reversed conical section. 
The principle of functioning is equal in all systems. The material suspended 
in a dense fluid (made of the heavy material mixed to water) is introduced 
as the tangent through the inlet, producing into the hydrocyclone a vortical 
movement. 
\par
Under the effect of centrifugal forces the more heavy mineral particles 
put into the dense means will try to go to the wall of the hydrocyclone. 
In the central part along the central axis, the vortex gives the origin to 
a strong depression that allows the presence of a central nucleus (air core). 
These particles will travel along spiral trajectories and will leave 
the separator through the apex (under the effect of gravitational forces) that 
is in the lower part. Instead, the lighter particles, 
following the most part of the fluid, 
will tend to go in the direction of the central part, always following 
a spiral trajectory where by means of a strong ascensional kinetic field 
(due to a convenient difference of pressure applied to the two exits) 
will leave the hydrocyclone through the vortex finder put in the upper part. 
\par
For the cylindrical hydrocyclone and for the multi--stage hydrocyclone
there is a tangential input of the dense means to originate a vortex 
and a central depression where an air core is 
situated. There will be a discharge in the lower part. The particles that are 
lighter will leave by a spiral trajectory through the vortex finder. 
The heavier particles will leave after going in the direction of 
the walls through the upper discharge (sink). 

\section{Confined swirling}

Because of the effect of the walls 
the flow in a hydrocyclone is the result of a movement of confined 
swirling (vortex) and recirculating flow, which are generated by 
the difference of the means. 
There are many applications of the movement with vortex: not only the 
solid--liquid separation considered here but also solid--gas, gas--liquid 
and gas--gas separation as in combustion, production of jet, stabilization 
of fluxes in a plasma, heat exchange, valves and turbomachines [8]. 
\par
The flux with vortex is the result of increasing the speed of 
the tangential component $v_t$ by using a tool called {\it vortex generator}. 
Vortices having enough force can produce in the inner part of the 
fluid mass some fluxes in the opposite direction of the vortex. 
The configuration of this recirculation zone can depend on many 
different factors as the distribution of velocity, the geometrical 
shape of the place where the vortex is developed, etc. 
\par
A certain number of properties characterize the movement 
with confined vortex: \\
a) the movement is 3--dimensional; \\
b) the tangential component $v_t$ of the velocity is predominant 
in the flux with the exception of the central turbulent part near the 
axis of the vortex; \\
c) the radial component $v_r$ of the velocity is very small; \\
d) the predominant component of 
the axial velocity $v_z$ is concentrated in a region close to the wall, 
having, close to the center, a component of inverse flux; \\
e) the axis of symmetry can become a spiral with a dynamical behaviour 
in time; \\
f) the microscopic flux is not very much related to the changes of velocity 
of the material into the system; \\
g) the turbulence is always very great because the flux with vortex is in 
a system of stationary type. 

\section{Properties of vortex movement}

Three are the properties that is necessary to introduce: \\
1) the Rankine profile for the tangential vorticity; \\
2) the vortex breakdown; \\
3) processing of vortex core. \\
As vortex of Rankine is intended the forced flux that can be interpreted as 
the combination of the rotation of a rigid body with contemporary 
presence of a free vortex where the action of the viscosity can be 
neglected. The rotation of the rigid body is classified 
as a rotational flux, instead the 
free vortex is classified as an irrotational flux. 
The zone where we have rotation of the rigid body is the central viscous 
region of the vortex and it is called nucleus of the vortex and also 
viscous nucleus. For a determined value of the axial coordinate such 
that the value taken by the tangential velocity is zero such position is 
the center of the vortex which usually is different from the center 
of the place where the vortex develops [9,10,11]. 
\par
To be able to deduce the characterized value of the Rankine vortex, 
one can follow the analysis prepared by Milne Thompson (1966). To such goal 
a plane flux is considered with a forced vortex of circular radius 
$r_b$ and constant vorticity ${\zeta}={\nabla} \wedge {v}$ 
externally limited by an irrotational vortex. 
\par
By cylindrical coordinates and avoiding the axial and azimuthal 
variation two circular trajectories are considered around 
the forced vortex and 
with radius $r_1<r_b$ and $r_b<r_2$ indicating by $v_{\phi_1}$ 
and $v_{\phi_2}$ the tangential velocity of the fluid in correspondence 
to the two limit trajectories. 
\par
Applying the Stokes theorem to the flux of the vector 
${\zeta}={ \nabla} \wedge {v}$ through the circular 
surfaces of ray $r_1$ and $r_2$ one obtains: 
\beq
\int_{S_i} ({\nabla} \wedge {v})\cdot {n} dS_i = 
\oint_{C_i}{v} dC_i \; \;\;\;\;\; i=1,2
\eeq
where ${n}$ indicates the versor of the normal to the generic surface 
$dS_i$ and $dC_i$ indicates an infinitesimal portion of the trajectory 
$C_i$ of length $2\pi r_i$. By the development of these formulas 
one obtains the circulations ${\Gamma}_1$ and ${\Gamma}_2$ 
\beq
{\Gamma}_i = ({\nabla} \wedge {v}) \pi r_i^2 = 
2\pi r_i {v}_{\phi_i} \; , \;\;\;\;\; i=1,2
\eeq
from which it is possible to obtain the tangential velocity for the 
Rankine vortex
\beq
{ v}_{\phi_i}={{\Gamma}_i \over 2\pi r_i} = 
{1\over 2} ({\nabla} \wedge {v}) r_i \; , \;\;\;\;\; i=1,2
\eeq
Remembering that the movement is realized into the plane $(r,\phi )$ the 
relation $2{\omega} = \nabla \wedge {v}$ applied to the 
direction $z$ gives 
\beq
v_{\phi_1} ={{\Gamma}_1\over 2\pi r_1} = {\omega} r_1 \; ,
\eeq
\beq
v_{\phi_2} ={{\Gamma}_2\over 2\pi r_2} = 
{\omega} {r_b^2\over r_2} \; .
\eeq
Therefore the forced vortex is equivalent to a rotation of a rigid body with 
angular velocity $\omega$. The tangential velocity $v_{\phi}(r)$ represents 
in the cartesian plane a line through the origin and of angular 
coefficient $\omega$ with $k_1=\omega$ and $k_2=\omega r_b^2$. 
The tangential velocity for forced vortex becomes 
$v_{\phi_1}={{\Gamma_1}\over 2\pi r_1}=k_1 r_1$ 
and for the free vortex $v_{\phi_2}={\Gamma_2\over 2\pi r_2}=k_2 /r_2$.  

\section{Break of the vortex}

If the tangential component of the velocity, 
namely of the Reynolds number, 
produces a zone of centered recirculation which signs the transition 
from a condition of flux with vortex that is said to be the break 
of the vortex and depend strongly from the swirl number $S$. This last one 
is a parameter that indicates the measure of the tangential velocity 
at the input. 
For two cylinders of ray $R_1$ and $R_2$ respectively that delimitate 
the zone of vorticity one has a $S$ with the following 
expression
\beq
S={ \int_{R_1}^{R_2} \rho r^2 v_z v_r dr \over
R_2 \int_{R_1}^{R_2} \rho r v_z^2 dr} \; . 
\eeq
Sarbkaya, Escudier and Zehuder have studied the problem of break of the 
vortex. All these studies make a relation between the Reynolds number 
and the circulation number $S$. 

\section{Precession of the nucleus of the vortex}

The recirculation zone generated by the break of the vortex is made by 
a jet with vortex in the strait direction and by the zone of inverse flux.  
These two quantities have distinct values for the axial and angular fluxes, 
sufficiently stable then exists an intermediate zone with great 
turbulence and fluctuations. 
This fluctuation originates the precession of the nucleus of forced 
vortex that tends to instability and moves around its symmetry axis. 
The different modes of precession of the nucleus and different stabilities 
are generally characterized using the Rayleigh number for the rotating fluid. 
This criteria says that a flux is stable if the quantity
\beq
\theta = \rho v_{\phi} r 
\eeq
grows with $r$ (forced vortex) and this quantity is stable if it is 
constant with $r$ (free vortex) and unstable if it decreases with $r$. 

\section{The nucleus of air}

If the energy of a movement with vortex is big enough one has the 
generation along the rotation axis of the fluid of a nucleus of air, 
where the transversal dimension depends directly on the pressure 
applied to the system and the form of such nucleus is cylindrical only 
in few cases. As it was said before, the axis of rotation does not coincide 
with the geometrical axis of the hydrocyclone. 
Namely for the nucleus of air there is no axial symmetry. 
Through the surface of the nucleus of air that is a liquid--air interphase 
there is a discontinuity of the gradient of the radial pressure due 
to the presence of two fluids with different density. 
The generation of the nucleus of air is due to the lowering of the pressure 
along the center of the vortex, to simplify the fluid is considered 
non viscous as was done by Milne Thompson. Considering 
a Rankine vortex one can indicate 
with $p_1$ and $p_2$ respectively the pressure into the vortex region, the 
pressure $p_1$ with the formula 
\beq
p_1 ={1\over 2}\rho \omega^2 r^2 + p_0 \; ,
\eeq
and for the potential vortex the application of Bernulli theorem gives
\beq
{p_2\over \rho} +{1\over 2} v_0^2 = {p_{\infty}\over \rho} \; ,
\eeq
where $v_{\phi}=0$ $p=p_{\infty}$ when $r$ goes to infinity. 
If the pressures are equal on the surface $r=r_b$ one obtains for $p_0$
\beq
p_0 = p_{\infty} - \rho \omega^2 r_b^2 \; ,
\eeq
\beq
p_1 = p_{\infty} - \rho \omega^2 r_b^2 (1 -{r^2 \over 2r_b^2}) \; ,
\eeq
\beq
p_2 = p_{\infty} - {1\over 2}\rho \omega^2 {r_b^2\over r^2} \; ,
\eeq
dividing by $p_{\infty}$ the last expression and introducing the quantity
\beq
A=\rho \omega^2 {r_b^2\over r_{\infty}} \; ,
\eeq
one obtains 
\beq
{p_1\over p_0} = 1 - A(1-{1\over 2}{r^2 \over r_b^2}) \; ,
\eeq
\beq
{p_2\over p_0} = 1 - {1\over 2} A {r^2 \over r_b^2} \; .
\eeq
The variation of the pressure in the two regions of the Rankine vortex gives a 
function of the quantity $r/r_0$ and a behaviour of parabolic type 
given by
\beq
{ p_1 \over p_{\infty} } - (1- A) = {1\over 2} A {r^2\over r_b^2} \; ,
\eeq
\beq
( {p_2 \over p_{\infty}} - 1) {r^2\over r_0^2} = {1\over 2} A \; .
\eeq
Different formulas have been given for the radius of the nucleus of air, 
the most important is that given by Concha and Barientos as
\beq
R_a={\sigma \over 2\mu \alpha - \Delta p_a} \; ,
\eeq
where $\alpha$ and $\Delta p$ depend on the geometrical 
parameters and $\mu$ is the viscosity of the fluid, $\sigma$ 
the superficial tension, $\alpha $ the radial component of the gradient 
of velocity evaluated in relation with $R_a$. $\Delta p$ is the difference 
of pressure through the interphase air--liquid. 

\section{The dynamical model}

The model of the Hydrocyclone is 3--dimensional, when to the sink and to 
the vortex finder a difference in the pressure is applied one 
has the generation of the air core where it is stable. 
The equations are the Navier--Stokes for the conservation of the mass and of 
the momentum [8]. We use the finite element method (FEM) [12] 
and the $k$--$\epsilon$ model [13], 
indicating by $(u_i,p,T,c,k,\epsilon )$ all the variables 
and taking into account that the $(k, \epsilon )$ turbulence model 
the field, this is characterized in terms of two variables using the 
turbulent energy $k$ which is defined as
\beq
k={1\over 2} u_i\cdot u_j \; ,
\eeq
\beq
\epsilon = v u_i \cdot u_j \; .
\eeq
In the following we will use repeated indices to indicate summation 
and the symbol $a_{,j}$ means ${\partial a\over \partial x_j}$. 
Typical turbulent eddy velocity and length scales (denoted by $u_i$ and 
$l_i$) can be characterized as $\sqrt{k}$ and $k^{1.5}/\epsilon$. 
$\mu_t =\rho_0 c_{\mu}k/\epsilon^2$ 
is the turbulent viscosity and is directly related 
to the turbulent quantities $k$ and $\epsilon$. 
A transport equation for $k$ can be obtained 
from the Navier--Stokes equations by a sequence of algebraic manipulations. 
This transport equation contains a number of unknown correlations. 
A second transport equation for $\epsilon$ can also be derived from 
the Navier--Stokes equations. Application of a number of modeling 
assumptions simplifies  these two equations to the well known equations 
of turbulent kinetic energy  and viscous dissipation of the $k$--$\epsilon$ 
model. Then the unknown variables become $(u_i,p,T,c,k,\epsilon )$  
and the corresponding field equations are: 
\beq
u_{j,j}=0 \; ,
\eeq
\beq
\rho_0 \big( {\partial u_i\over \partial t} + u_ju_{i,j}  \big) 
= - p_{,i}+\rho_0 f_i - \rho_0 g_i [\beta_T (T-T_0) -\beta_c c] 
+ [\mu (u_{i,j} + u_{j,i})_{,j} ] \; ,
\eeq
\beq
\rho_0 c_p \big( {\partial T\over \partial t} + u_jT_{,j}  \big) 
= (\lambda T_{,j})_{,j}+\mu \Phi + H \; ,
\eeq
\beq
\rho_0 \big( {\partial c\over \partial t} + u_jc_{,j}  \big) 
= \rho_0 (\alpha c_{,j})_{,j} + q_c + R \; ,
\eeq
\beq
\rho_0 \big( {\partial k \over \partial t} + u_jk_{,j} \big) 
= \big( \mu_0 +{\mu_t\over \sigma_{k}} k_{,j}\big)_{,j}
+ \mu_t \Phi + \mu_t g_i 
\big( {\beta_t\over \sigma_t} T_{,j} +{\beta_c\over S_t} c_{,j} \big)
-\rho_0 \epsilon  \; ,
\eeq
\beq
\rho_0 \big( {\partial \epsilon \over \partial t} + u_j\epsilon_{,j} \big) 
= \big( \mu_0 +{\mu_t\over \sigma_{\epsilon}} \epsilon_{,j}\big)_{,j}
+ c_1 {\epsilon \over k} \mu_t \Phi +c_1 (1-c_3){\epsilon \over k} g_j 
\big( {\mu_t\over \sigma_t} \beta_T T_{,j} +{\mu_t\over S_t}\beta_c c_{,j}\big)
-\rho_0 c_2 {\epsilon^2 \over k} \; ,
\eeq
where $\Phi =2 \epsilon_{ij}\epsilon_{ij}$, with 
$\epsilon_{ij}={1\over 2}(u_{i,j}+u_{j,i})$ the strain stress tensor.
$\alpha$ is the slip coefficient, $\beta_T$ and $\beta_c$ are 
the volume expansion coefficients, $\lambda$ is the thermal 
conductivity, $g_i$ is the gravitational force vector. 
After years of testing the $k$--$\epsilon$ model, 
the choice of the quantity
taken as empirical constants has led to the following recommended 
set of model constants: $c_{\mu}=0.09$, $\sigma_k = 1.00$, 
$\sigma_{\epsilon} =1.30$, $\sigma_{\tau} = 0.9$, $S_t=0.9$, 
$c_1 =1.44$, $c_2=1.92$, $c_3=0.8$. 

\section{Turbulence Modelling} 

The system of equations (21--26) is discretized by the usual finite 
element method. It is important to note that the introduction of 
the eddy viscosity concept makes the total effective viscosity a function  
of position, thus necessitating a stress divergence formulation.  
Using the approximation 
\beq
k({x},t)=\phi^T {K}(t) \; ,
\eeq
\beq
\epsilon ({x}, t) = \phi^T {E}(t) \; .
\eeq
Within each element, 
the velocity, pressure and temperature field are approximated by
\beq
u_i({x},t)=\phi^T { U}_i (t) \; , 
\eeq
\beq
p({x},t)=\psi^T {P} (t) \; ,
\eeq
\beq
T({x},t)=\theta^T {T} (t) \; .
\eeq
where ${U}_i$, ${P}$ and ${T}$ are column vectors of element 
nodal point unknowns and $\phi$, $\psi$ and $\theta$ are column vectors of 
the interpolation functions. Herein the same basis functions are employed 
but cost--effective restrictions. Substitution of these approximations 
into the field equations and boundary conditions yields a set of equations: 
\beq
{f}_1 (\phi , \psi , \theta , {U}_i , 
{P}, {T}) = {R}_1 \; ,
\eeq
\beq
{f}_2 (\phi , {U}_i ) = {R}_2 \; ,
\eeq
\beq
{f}_3 (\phi , \theta , {U}_i , {T}) = {R}_3 \; ,
\eeq
where ${R}_1$, ${R}_2$, and ${R}_3$ are the residuals 
(errors) resulting from the use of the approximations of equation (3).
The process results in the following finite system of nonlinear ordinary 
differential equations: 
$$
\left( \begin{array}{ccccccc}  
         M & 0 & 0 & 0 & 0 & 0 & 0 \\ 
         0 & M & 0 & 0 & 0 & 0 & 0 \\ 
         0 & 0 & M & 0 & 0 & 0 & 0 \\ 
         0 & 0 & 0 & N & 0 & 0 & 0 \\ 
         0 & 0 & 0 & 0 & 0 & 0 & 0 \\ 
         0 & 0 & 0 & 0 & 0 & M & 0 \\ 
         0 & 0 & 0 & 0 & 0 & 0 & M  
\end{array}  \right) 
\left( \begin{array}{c}  
       {\dot U}_1 \\{\dot U}_2 \\{\dot U}_3 \\{\dot T} \\{\dot P} 
\\{\dot K}_1 \\{\dot E}_1 
\end{array}  \right) +
$$
\beq
+ \left( \begin{array}{ccccccc}  
         K_{11} & K_{12} & K_{13} & K_{14} & -C_1 & 0 & 0 \\ 
         K_{21} & K_{22} & K_{23} & K_{24} & -C_2 & 0 & 0 \\ 
         K_{31} & K_{32} & K_{33} & K_{34} & -C_3 & 0 & 0 \\ 
         0 & 0 & 0 & K_{44} & 0 & 0 & 0 \\ 
   -C_1^T & -C_2^T & -C_3^T & 0 & 0 & 0 & 0 \\ 
         0 & 0 & 0 & 0 & 0 & K_{66} & 0 \\ 
         0 & 0 & 0 & 0 & 0 & 0 & K_{77} 
\end{array}  \right) 
\left( \begin{array}{c}  
       {U}_1 \\{U}_2 \\{U}_3 \\{T} \\{P} \\K \\E 
\end{array}  \right) = 
\left( \begin{array}{c}  
       F_1 \\F_2 \\F_3 \\F_4 \\0 \\F_6 \\F_7 
\end{array}  \right) \; .
\eeq
The sub-matrices $M$, $N$ and $C_i$ are given by
\beq
M=\int_V \rho_0 \phi \phi^T dV \; ,
\eeq
\beq
N=\int_V \rho_0 c_p \theta \theta^T dV \; ,
\eeq
\beq
C_i= \int {\partial \phi \over \partial x_j} \psi^T dV \; .
\eeq
The remaining sub-matrices take the following form 
for $i,j=1,2,3$. Note that parentheses on repeated indices indicate a 
relaxation of the summation convention: 
\beq
K_{(i)(i)} = \int_V \rho_0 u_j \phi\phi_{,j}^T dV 
+ \int_V \mu (\phi_{,j}\phi_{,j}^T + \phi_{,(i)}\phi_{,(i)}^T ) dV \; ,
\eeq
\beq
K_{ij}= \int_V \mu \phi_{,i} \phi_{,j}^T dV \;\;\;\;\; i\neq j \; ,
\eeq
\beq
K_{i4}=\int_V \rho_0 \beta g_i \phi \phi^T dV \; ,
\eeq
\beq
K_{44}=\int_V \rho_0 c_p u_i \phi \phi_{,j}^T dV + 
\int_V \lambda \phi_{,j}\phi_{,j}^T dV \; ,
\eeq
\beq
K_{66}=\int_V \rho_0 u_j \phi \phi_{,j}^T dV 
+ \int_V {\mu_t \over \sigma_{\epsilon}} \phi_{,j}\phi_{,j}^T dV 
+ \int_V \rho_0 {\epsilon \over k} \phi \phi^T dV \; ,
\eeq
\beq
K_{77}=\int_V \rho_0 u_j \phi \phi_{,j}^T dV + 
\int_V {\mu_t\over \sigma_{\epsilon}}\phi_{,j} \phi_{,j}^T dV 
+\int_V \rho c_2 {\epsilon \over k} \phi \phi^T dV \; ,
\eeq
\beq
F_{i}= 
\int_V \rho f_i dV + \int_V \rho g_i (1+\beta T_0) \phi dV \; ,
\eeq
\beq
F_4= - \int_S \lambda \theta {\partial T\over \partial n} dS + 
\int_V qs \theta dV + \int_V \mu \Phi \theta dV \; ,
\eeq
\beq
F_6= \int_S {\mu_t \over \sigma_k} {\partial k\over \partial n} \phi
dS + \int_V \mu_t \Phi \phi dV \; ,
\eeq
\beq
F_7= \int_S {\mu_t\over \sigma_{\epsilon}} {\partial \epsilon\over
\partial n} \phi dS + \int_V c_1 {\epsilon \over k} \mu_t \Phi \phi dV
\; .
\eeq
Each of the above integrals are evaluated using the isoparametric 
map/quadrature procedure. Then, using the isothermal hypothesis, 
the solution is obtained by applying 
the Galerkin Finite Element Method in the usual fashion. 
In Figures 1--3 we show our FEM numerical results, which are in good 
agreement with the experimental data [5,6,7]. See figure captions 
for further details. 

\section{A simple model for the Hydrocyclone}

In this section we study a simple 3D 
family of maps which shows some of the properties of the Hydrocyclone. 
\par
We call $(\rho ,\theta ,v_z)$--polypous
the family of planar maps $f_{\rho \theta}:{\bf R}^3\to {\bf R}^3$ such that
$f_{\rho \theta v_z}(x,y,z)=(f^{(1)}_{\rho \theta v_z}(x,y,z),
f^{(2)}_{\rho \theta v_z}(x,y,z),f^{(3)}_{\rho \theta v_z}(x,y,z))$, 
with $\rho \geq 0$, $\theta \in [-2\pi ,2\pi ]$, $v_z\in {\bf R}$ and:
$$
f_{\rho \theta v_z}^{(1)}(x,y,z)= 
x \; \rho \cos{\theta} + y \; \rho \sin{\theta} -
{x^3\over 3} -{y^3\over 3} \; ,
$$
\beq
f_{\rho \theta v_z}^{(2)}(x,y,z)= 
- x \; \rho \sin{\theta} + y \; \rho \cos{\theta}
+ {x^3\over 3} - {y^3\over 3} \; ,
\eeq
$$
f_{\rho \theta v_z}^{(3)}(x,y,z)= z + v_z \; .
$$
In Figure 4 and 5 we show the dynamics of our map for $\rho=1$, 
$v_z=-1$ and three different values of the $\theta$ parameter. 
The trajectories follow a spiral and tend to go down in the direction 
of the central part. By taking a positive value 
for the velocity $v_z$ and different values
for the parameter $\theta$ we can model also the 
trajectories of particles which go in the upper part.  
\par 
Because the dynamics of the $(\rho ,\theta , v_z)$--polypous in the 
$z$ direction is trivial, we analyze in detail the 
$(\rho ,\theta ,v_z)$--polypous on the $(x,y)$ plane for $v_z=0$ 
(see also [14]). 
\par
To characterize the $(\rho ,\theta )$--polypous we can calculate the Jacobian
\beq
Df_{\rho \theta}(x,y)=
\left( \begin{array}{cc}
 \rho \cos{\theta} - x^2 & \rho \sin{\theta} - y^2 \\
-\rho \sin{\theta} + x^2 & \rho \cos{\theta} - y^2
\end{array} \right) \; .
\eeq
The trace of $Df_{\rho \theta}$ is given by
\beq
Tr(Df_{\rho \theta}(x,y))=2\rho \cos{\theta} -(x^2+y^2) \; ,
\eeq
and the determinant
\beq
det(Df_{\rho \theta}(x,y))=2 x^2 y^2 -\rho (\cos{\theta} + \sin{\theta})
(x^2 +y^2) + \rho^2 \;.
\eeq

A direct calculus shows that the eigenvalues of
the $\rho \theta$--polypous are given by
\beq
\lambda_{1,2}(x,y)={1\over 2}\big[ Tr(Df_{\rho \theta}(x,y)) \mp
\sqrt{Tr(Df_{\rho \theta}(x,y))^2 - 4 det(Df_{\rho \theta}(x,y))} \big] \; .
\eeq

\par
The $(\rho ,\theta )$--polypous has a fixed point in
$(0,0)\in {\bf R}^2$.
This fixed point is hyperbolic if $\rho \neq 1$. It is asymptotically
stable, i.e an attractor, if $\rho < 1$ and it is unstable if $\rho >
1$. The $(\rho ,\theta )$--polypous is differentiable but it is not a
homeomorphism because it is not invertible: there are several anti--images
of the origin, one of them is obviously the origin itself.

\vskip 0.5 truecm
\par
Now we discuss in great details the very interesting case $\rho =1$: 
The $(\rho ,\theta )$--polypous has a non--hyperbolic 
fixed point in $(0,0)\in {\bf R}^2$ if $\rho =1$. 
This fixed point is asymptotically stable 
if $\theta \in \; ]-{\pi /4},{3\pi /4}[$, but $\theta \neq 0$. 
First we observe that when $\theta =0$ and $\theta =\pi$ the eigenvalues 
of $Df_{1\theta}(0,0)$ are real numbers. For all the other values 
of $\theta$ the eigenvalues are complex but not real and have 
unitary modulus. 
\par
We call $d_{\theta}(x,y)=det(Df_{1\theta}(x,y))$ and obtain
$$
\nabla d_{\theta}(x,y)= ( 4 x y^2 - 2 (\cos{\theta} + \sin{\theta})x ,
4 x^2 y - 2 (\cos{\theta} + \sin{\theta})y) ,
$$
$$
Hd_{\theta}(x,y)=
\left( \begin{array}{cc}
4 y^2 - 2 (\cos{\theta} + \sin{\theta}) &  8 x y \\
8 x y & 4 x^2 - 2 (\cos{\theta} + \sin{\theta})
\end{array} \right) \; , 
$$
and its eigenvalues in (0,0) are equal to 
$- 2 (\cos{\theta} + \sin{\theta})$. The Hessian $Hd_{\theta}(0,0)$ 
is negative definite if $(\cos{\theta} + \sin{\theta}) > 0$, thus if 
$\theta \in ]-{\pi /4},{3\pi /4}[$. For these values of $\theta$ 
the origin is a local maximum and by using the theorem of reference 
[14] ($\theta \neq 0$) we have that $(0,0)$ is asymptotically stable. 
\par
Consider $\theta = 2\pi /n$, $n\geq 3$. The dynamics of $f_{1 \theta}$ is 
attractive and, because $Df_{1\theta}(0,0)$ is a rotation of $\theta$, 
it looks like a polypous with $n$ branches. 
\par
Moreover, if we add to the first two components 
of the $(\rho ,\theta ,v_z)$--polypous an 
oscillatory term $\cos{(\omega z)}$ we can get an elicoidal cylinder 
superimposed to the spiral dynamics (see Figure 6). 
So the modified $(\rho ,\theta ,v_z)$--polypous give rise also to 
the elicoidal effect, which has been recently observed experimentally [7]. 

\section{Conclusions}

In this paper we have studied the turbulence and bifurcation 
in the motion of an Hydrocyclone by using a Finite Element Method 
based on the Navier--Stokes equations. 
The numerical results are in good agreement with the experimental 
data. We have also analyzed a simple 3D family of maps which 
models very well the behaviour of the trajectories of the light and 
heavy particles in the Hydrocyclone. 

\section*{References}

\begin{description}

\item{\ [1]} J.L. Smith, ``An analysis of the vortex
flow in the cyclone separator'', Jour. of Basic Engineering--Trans.
of ASME, 609 (1962).

\item{\ [2]} D. Bradley, {\it The hydrocyclones} (Pergamon, London,
1965).

\item{\ [3]} J. Mizrahi and E. Cohen, ``Studies of some factor
influencing the action of hydrocyclones'', Trans. Instr. Min. Metall.
{\bf 75}, C318 (1966).

\item{\ [4]} M.I.G. Bloor and D.B. Ingham,
``The flow in industrial cyclones'', Jour. of Fluid Mechanics,
{\bf 178}, 507 (1987).

\item{\ [5]} A. Barrientos, R. Sampaio 
and F. Concha, "Effect of air core on the 
preformation of a Hydrocyclone", in Proceedings XVIII International 
Mineral Processing Congress, May 1993, Sidney, pp. 267--270.

\item{\ [6]} F. Concha, A. Barrientos,
J. Montero and R. Sampaio, ``Air core and roping in hydrocyclones'',
Symposium on Comminution 1994, Stockolm, pp. 814--823.

\item{\ [7]} B. Chine, F. Concha and A.
Barrientos, ``A finite difference solution of the swirling flow
in a hydrocyclone'', in Proc. of International Conference on
Finite Elements in Fluids -- New Trends and Applications,
Venezia, 15--21 October, 1995.

\item{\ [8]} G.K. Batchelor, 
{\it An introduction to fluid dynamics} (Cambridge University 
Press, Cambridge, 1967).

\item{\ [9]} S.V. Patankar and D.B. Spalding, ``A calculation
procedure for heat, mass and momentum transfer in three--dimensional parabolic
flows'',
Int. Jour. of Heat and Mass Transfer, {\bf 15}, 1787 (1972).

\item{\ [10]} S.V. Patankar, {\it Numerical
heat transfer and fluid flow} (Hemisphere Publ. Corp., New York,
1980).

\item{\ [11]} K.T. Hsieh and K. Rajamani, 
"Mathematical model of the hydrocyclone 
based on physics of fluid flow" AIChE Jour. {\bf 37}, 735 (1991).

\item{\ [12]} G.F. Carey and J.T. Oden, {\it Finite Elements: 
Fluid Mechanics} (Prentice--Hall, New Jersey, 1986). 

\item{\ [13]} Fluid Dynamics Analysis Package, FIDAP 7.0, 
Fluid Dynamics International, Inc. (1993). 

\item{\ [14]} M. Morandi Cecchi and L. Salasnich, 
``Non-hyperbolic dynamics: a family of special functions'', 
Open System and Information Dynamics, vol. 5 (1998). 

\end{description}

\newpage

\section*{Figure Captions}

\vskip 0.5 truecm

\begin{description}

\item{\ {\bf Figure 1}}: FEM trajectories of heavy 
particles in the Hydrocyclone. 
\vskip 0.4 truecm
\item{\ {\bf Figure 2}}: FEM trajectories of light 
particles in the Hydrocyclone. 
\vskip 0.4 truecm
\item{\ {\bf Figure 3}}: FEM velocity field in the $z$ direction 
for the Hydrocyclone. 
\vskip 0.4 truecm
\item{\ {\bf Figure 4}}: Spiral trajectories obtained with 
the $(1,\theta,-1)$--polypous in the $(x,y)$ plane. 1) 
$\theta =0.1 \cdot \pi$, 
2) $\theta  =0.03 \cdot \pi $, 3) $\theta =0.05 \cdot \pi $. 
\vskip 0.4 truecm
\item{\ {\bf Figure 5}}: Same trajectories of Figure 4 in 3D. 
\vskip 0.4 truecm 
\item{\ {\bf Figure 6}}: Elicoidal cylinder 
obtained with the modified $(1,0.2 \cdot \pi,-1)$--polypous 
with $\omega = 0.2$. 

\end{description}

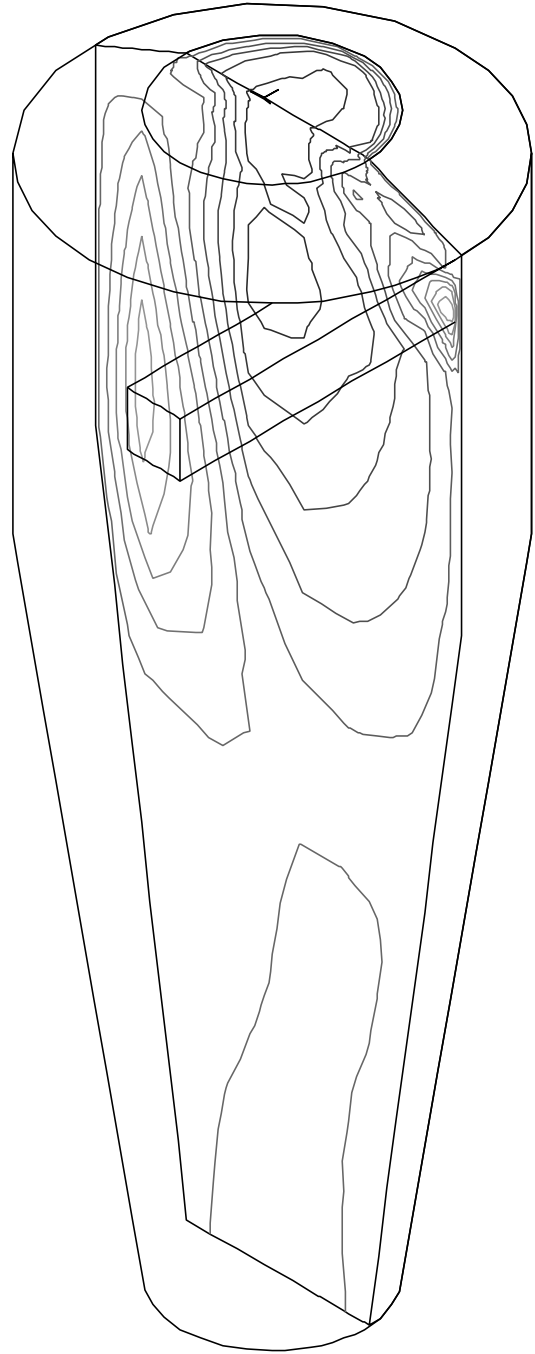
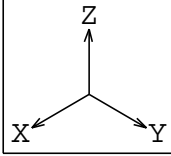
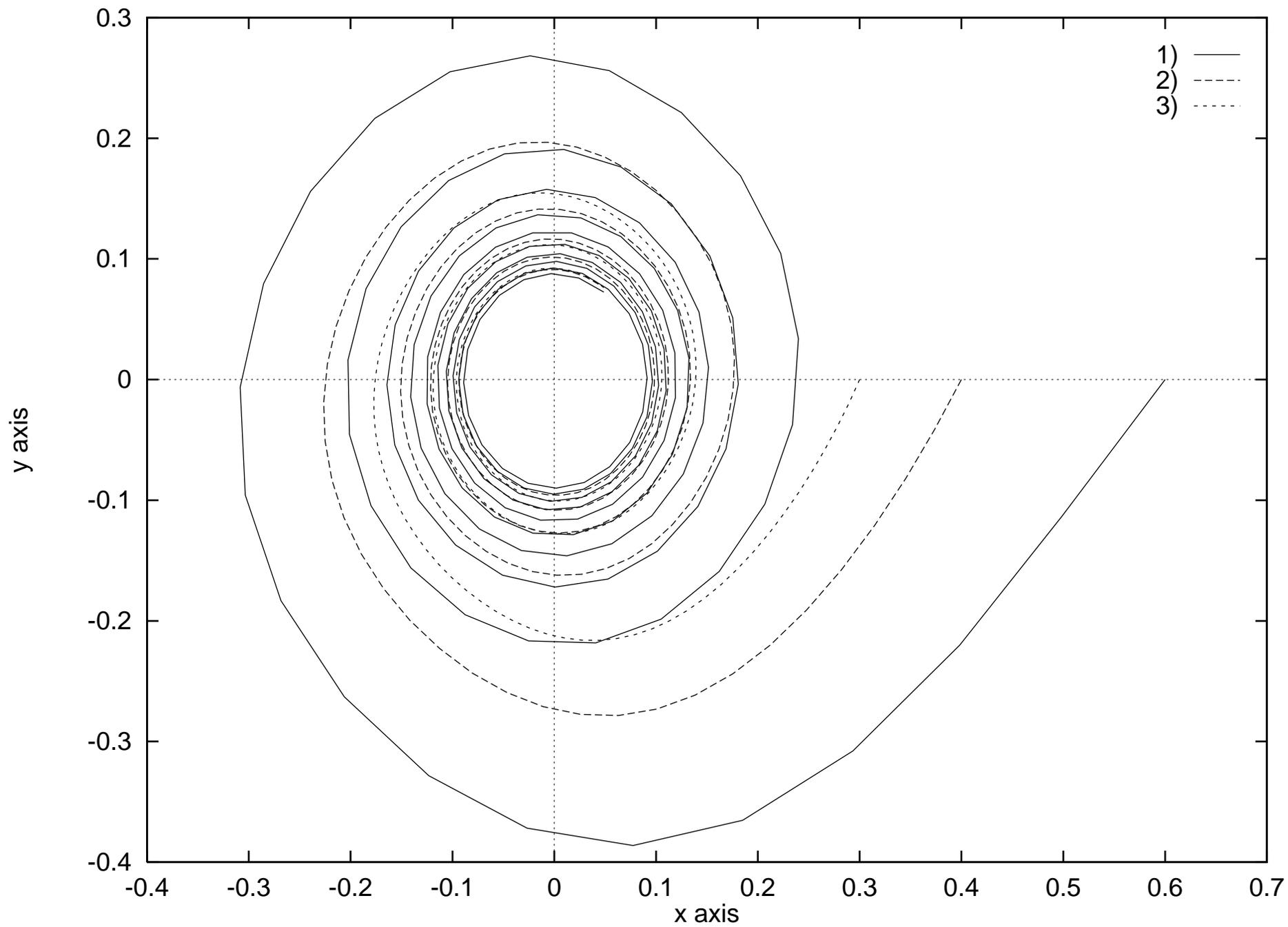
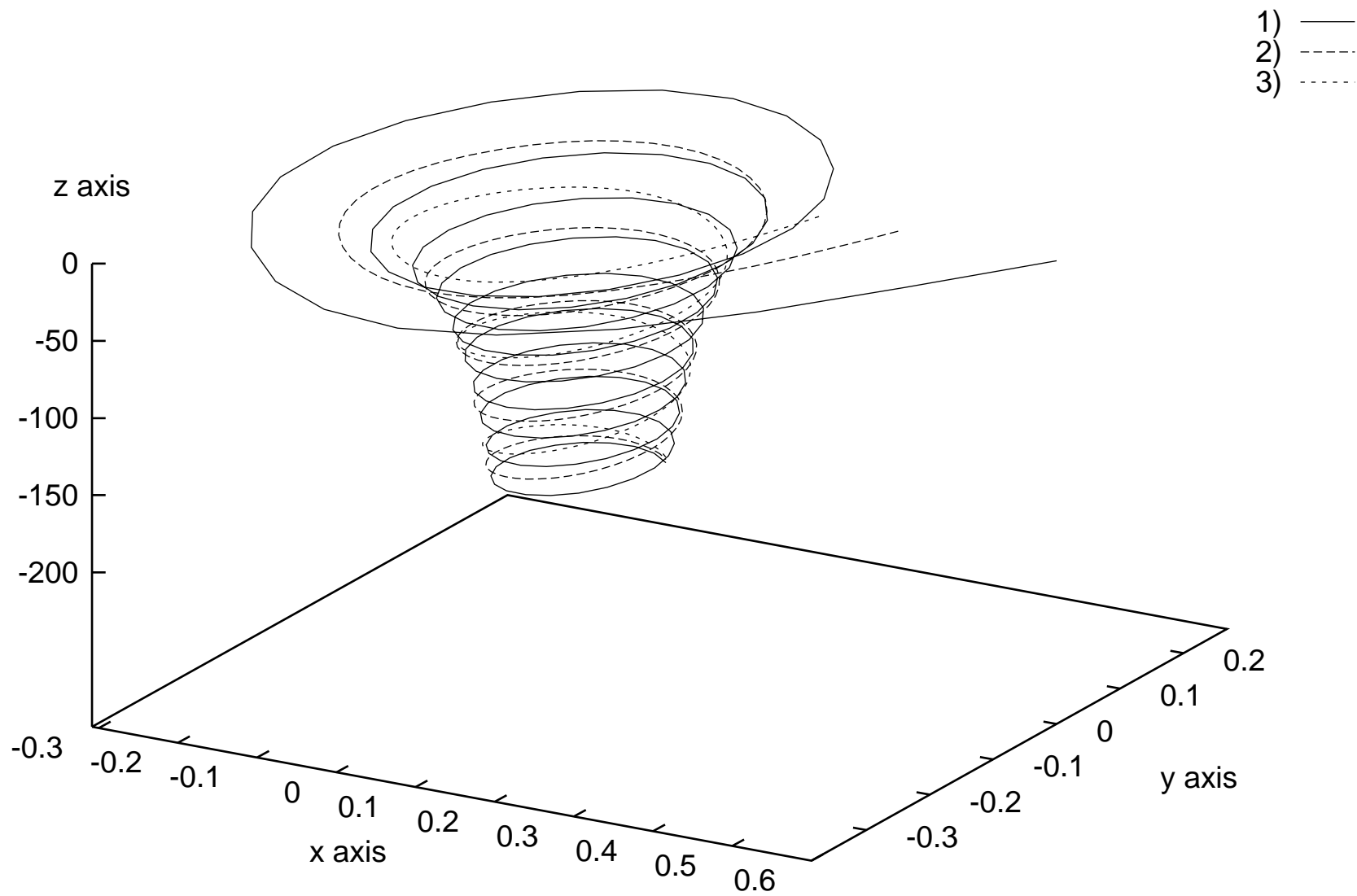
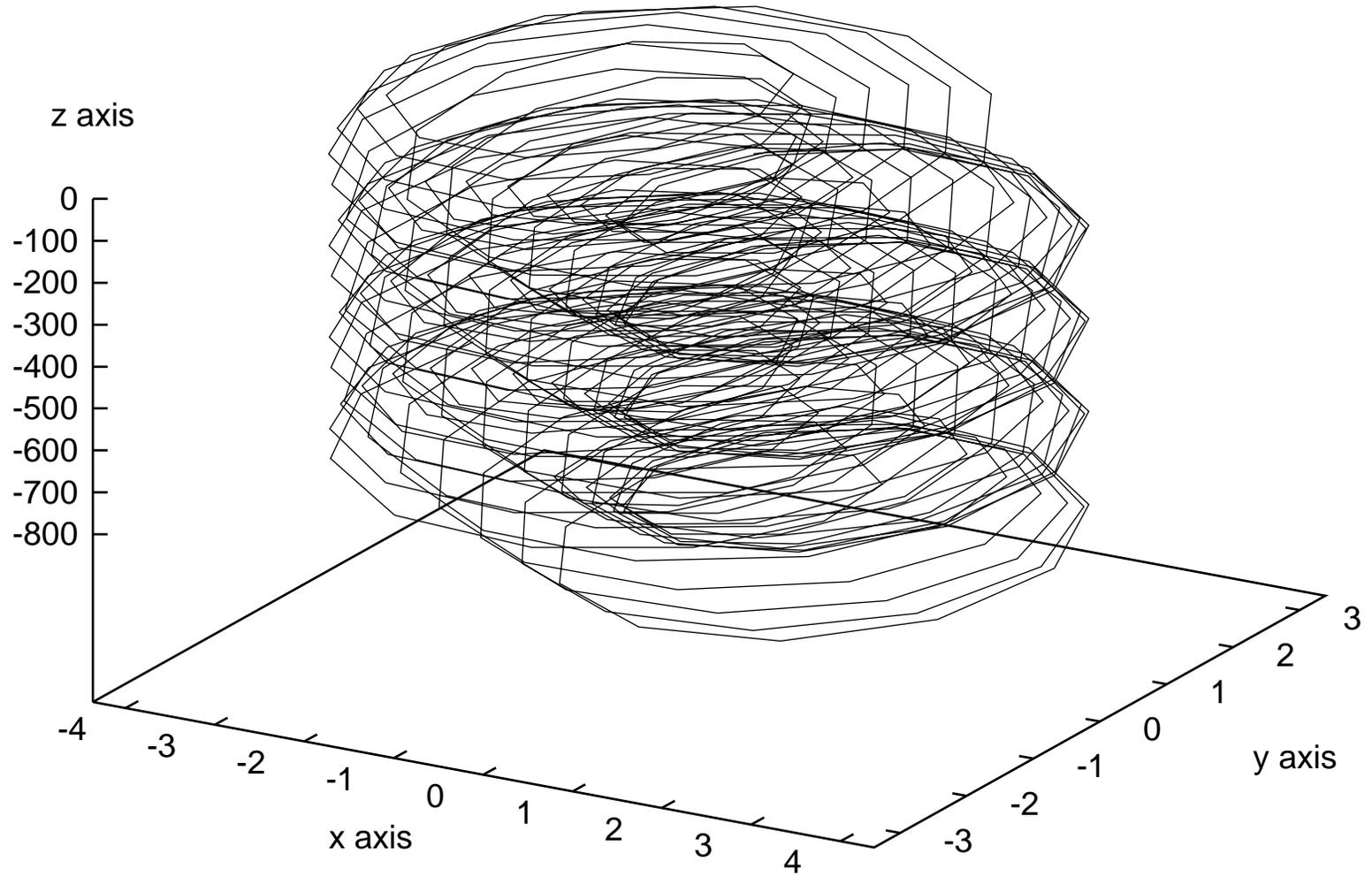
\end{document}